\documentclass[12pt]{article}
\usepackage{times}
\usepackage{geometry}
\usepackage{graphicx}
\usepackage[dvipsnames]{xcolor}
\usepackage{natbib}
\usepackage{enumitem,amssymb}
\usepackage{ragged2e}
\newlist{thematic}{itemize}{8}
\setlist[thematic]{label=$\square$}
\usepackage{pifont}

\usepackage{hyperref}
\newcommand\myshade{100}
\colorlet{mylinkcolor}{violet}
\colorlet{mycitecolor}{YellowOrange}
\colorlet{myurlcolor}{Aquamarine}
\hypersetup{
  linkcolor  = mylinkcolor!\myshade!black,
  citecolor  = mycitecolor!\myshade!black,
  urlcolor   = myurlcolor!\myshade!black,
  colorlinks = true,
}
\usepackage[capitalize]{cleveref}

\begin{document}

\title{International Astrophysical Consortium for High-energy Calibration
\\
\bf{Summary of the 15th IACHEC Workshop}
}

\author{
K.~K.~Madsen$^1$,
V.~Burwitz$^2$,
K.~Forster$^3$
C.~E.~Grant$^4$,
M.~Guainazzi$^5$, \and
V.~Kashyap$^6$, 
H.~L.~Marshall$^4$, 
E.~D.~Miller$^4$, 
L.~Natalucci$^7$, 
P.~P.~Plucinsky$^6$, \and
Y.~Terada$^{8,9}$
}

\date{\today}   

\newcommand{\artxc}{\textit{ART-XC}}
\newcommand{\erosita}{\textit{eROSITA}}
\newcommand{\rosat}{\textit{ROSAT}}
\newcommand{\xmm}{XMM-\textit{Newton}}
\newcommand{\chandra}{\textit{Chandra}}
\newcommand{\suzaku}{\textit{Suzaku}}
\newcommand{\swift}{\textit{Swift}}
\newcommand{\nicer}{\textit{NICER}}
\newcommand{\astrosat}{\textit{AstroSat}}
\newcommand{\nustar}{\textit{NuSTAR}}
\newcommand{\hxmt}{\textit{Insight-HXMT}}
\newcommand{\hitomi}{\textit{Hitomi}}
\newcommand{\xrism}{\textit{XRISM}}
\newcommand{\integral}{\textit{INTEGRAL}}
\newcommand{\fermi}{\textit{Fermi}}
\newcommand{\athena}{\textit{Athena}}
\newcommand{\ep}{\textit{Einstein Probe}}
\newcommand{\ixpe}{\textit{IXPE}}
\newcommand{\leia}{\textit{LEIA}}
\DeclareRobustCommand{\ion}[2]{\textup{#1\,\textsc{\lowercase{#2}}}}
\newcommand{\cstat}{{\tt c-stat}}

\maketitle

{\centering
  $^1$NASA Goddard Space Flight Center, USA \\
  $^2$Max Planck Institute for Extraterrestrial Physics, Germany\\
  $^3$Cahill Center for Astronomy and Astrophysics, California Institute of Technology, USA \\
  $^4$Kavli Institute for Astrophysics and Space Research, Massachusetts Institute of Technology, USA \\
  $^5$ESA-ESTEC, The Netherlands \\
  $^6$Center for Astrophysics $|$ Harvard \& Smithsonian (CfA), USA \\
  $^7$IAPS-INAF, Italy \\
  $^8$Saitama University, Japan \\
  $^9$Japan Aerospace Exploration Agency, Institute of Space and Astronautical Science, Japan
}

\vspace{5mm}
\begin{center}
{\bf \large Abstract} \\
\end{center}

In this report we summarize the activities of the International Astronomical Consortium for High Energy Calibration (IACHEC) from the 15th IACHEC Workshop in Pelham, Germany. Sixty scientists directly involved in the calibration of operational and future high-energy missions gathered during 3.5 days to discuss the status of the cross-calibration between the current international complement of X-ray observatories, and the possibilities to improve it. This summary consists of reports from the Working Groups with topics ranging across: the identification and characterization of standard calibration sources, multi-observatory cross-calibration campaigns, appropriate and new statistical techniques, calibration of instruments and characterization of background, preservation of knowledge, and results for the benefit of the astronomical community.

\section{Meeting Summary}

The International Astronomical Consortium for High Energy Calibration (IACHEC)\footnote{\url{http://iachec.org}} is a group dedicated to supporting the cross-calibration of the scientific payload of high energy astrophysics missions with the ultimate goal of maximizing their scientific return. Its members are drawn from instrument teams, international and national space agencies, and other scientists with an interest in calibration. Representatives of over a dozen current and future missions regularly contribute to the IACHEC activities. 

IACHEC members cooperate within Working Groups (WGs) to define calibration standards and procedures. The objective of these groups is primarily a practical one: a set of data and results produced from a coordinated and standardized analysis of high-energy reference sources that are in the end published in refereed journals. Past, present, and future high-energy missions can use these results as a calibration reference. Table \ref{table:WGoverview} summarizes the WGs active during this report and their primary projects and areas of responsibility.

The 15th IACHEC meeting was hosted by Vadim Burwitz (MPE, Garching, Germany). It was held at Seeblick,  Pelham in Germany, and the meeting was attended by 60 scientists representing high energy missions internationally. The meeting also featured presentations from IACHEC members unable to attend in person, who were able to view and contribute to the meeting via video teleconferencing. Advances in the understanding of the calibration of more than a dozen missions was discussed, covering multiple stages of operation and development:

\begin{itemize}
\item[-] Currently operating missions - \chandra, \erosita, \hxmt, \integral, \ixpe, \leia, \nicer, \nustar, \swift, \xmm
\item[-] Pre-launch status - \ep, \xrism
\item[-] Missions under development: \athena, \textit{COSI}, \textit{LEXI}, \textit{SMILE}, \textit{SVOM}

\end{itemize}

Alongside the contributed sessions, the IACHEC WGs held their meetings and this report summarizes the main results of the 15th meeting\footnote{The presentations made at the meeting are available at: \url{https://iachec.org/2022-2}}, and comprises the reports from each of the IACHEC WGs.

The IACHEC gratefully acknowledges sponsorship for meeting from AHEAD 2020 and the Society for Promotion of Space Science and the International Conference Support Program of Kanagawa Prefecture Japan.

\begin{table}[t]
    \centering
    \begin{tabular}{| p{0.30\linewidth}|p{0.2\linewidth}|p{0.5\linewidth}|}
        \hline
         Working Group & WG Chair & Projects  \\
         \hline
         \hline
         Calibration Statistics & Vinay Kashyap & Quantifying response uncertainties; statistical methods; Concordance project \\
         \hline
         Clusters of Galaxies & Eric Miller & Multi Mission Study of selected targets from the HiFLUGCS sample\\
         \hline
         Contamination & Herman Marshall & Definition, measurement, and mitigation of molecular contaminant \\
         \hline
         Coordinated Observations & Karl Forster & Organization of yearly cross-calibration campaign of 3c273; investigation of potential cross-calibration candidate 1ES 0229+200; publication of 3c273 cross-calibration campaign\\
         \hline
         Detectors and Background & Catherine Grant & Forum for discussion of detector effect; background modeling\\
         \hline
         Heritage & Matteo Guianazzi & Curation of the IACHEC work; the IACHEC source database (\href{http://iachecdb.iaps.inaf.it}{\tt ISD})\\
         \hline
         Non-thermal SNR & Lorenzo Natalucci & Cross-calibration with the Crab and G21.5-0.3\\
         \hline
         Thermal SNR & Paul Plucinsky & Cross-calibration with 1E 0102.2$-$7219 (E0102), N132D, and Cas~A; definition of standard models \\
         \hline
         Timing & Yukikatsu Terada & Summary of Timing performance and calibration status across missions; Systematic timing cross-calibration with Crab archive data; Planning timing cross-calibration campaign\\
         \hline
         White Dwarf and Isolated Neutron Stars & Vadim Burwitz & Cross-calibration with RX\,J1856.5$-$3754 and 1RXS\,J214303.7$+$065419 \\
         \hline
    \end{tabular}
    \caption{IACHEC Active Working Groups}
    \label{table:WGoverview}
\end{table}

\section{Working Group reports}

\subsection{Calibration Uncertainties: CalStats}\label{s:calstat}

The Calibration Statistics Working Group (CalStats WG\footnote{\url{https://iachec.org/calibration-statistics/}}) was set up as a forum for discussion of statistcal, methodological, and algorithmic issues that affect the calibration of instruments, how calibration data are used in data analysis, and how analysis results are interpreted.  Its purview includes documenting and recommending good analysis practices, and developing mathematically robust techniques to deal with commonly encountered calibration analysis problems.

During IACHEC\,15, CalStats WG talks spanned two sessions on April 25, 2023.  During the sessions, there were talks by: 
Herman Marshall on devising a better system for fitting polarization data utilizing the information in event charge deposition ellipticities (Marshall 2021, 2024);
Stefano Silvestri on describing the effects of systematic uncertainties in IXPE effective areas combined with modulation factors (Silvestri et al.\ 2023);
Benjamin Schneider on CCD events energy calibration via correlations of synthetic vs raw ADU spectra (Schneider et al.\ 2022);
Ivan Valtchanov on denoising and super-resolving XMM data by leveraging machine learning methods and training on simulations (Sweere et al.\ 2022); 
Craig Markwardt on the practical use of SCORPEON,\footnote{\url{https://heasarc.gsfc.nasa.gov/docs/nicer/analysis_threads/scorpeon-overview/}} the background model developed for NICER; and
Vinay Kashyap on the difficulties arising from counts sparsity in high-resolution spectra with weak sources and a new method to characterize how good a model for the background is (Zhang et al.\ 2023).

In addition, there were two panel discussions, one focusing on the Concordance project, led by Yang Chen and Herman Marshall (see Chen et al.\ 2019, Marshall et al.\ 2021), and one on \cstat\ and systematic effects in spectral analysis, led by Max Bonamente and Yang Chen (Bonamente 2020, 2022).

There were also several talks during other Working Group sessions which had elements of interest to the CalStats WG: Daniel Wik (on galaxy cluster temperature cross-telescope comparisons), Jeremy Sanders (on Chandra vs eROSITA flux comparisons), Konrad Dennerl (on modeling the arf/rmf of eROSITA), Jukka Neveleinen (on XMM-Chandra cluster-to-cluster scatter), Christian Pommranz (on the CORRAREA tool to match the effective area of XMM/EPIC-MOS to XMM/EPIC-pn), and Jelle de Plaa (on the SPEX analysis package).

The CalStats WG maintains a library of published papers of interest on various topics of interest to WG members on the IACHEC web site\footnote{\url{https://iachec.org/calibration-statistics/\#library}}.  
The WG also maintains the current best known information about background models for various instruments at the IACHEC wiki\footnote{\url{https://wikis.mit.edu/confluence/display/iachec/Calibration+Statistics}}.

Future plans of the WG include holding virtual talks of interest, and continuing work on projects ranging from calibration uncertainty, the Concordance Project, polarization fitting methods, superresolution and deconvolution using both Bayesian and machine learning methods, spatio-spectral modeling and disambiguation, principled background modeling, etc.

\subsection{Clusters of Galaxies}\label{s:clusters}

This WG uses clusters of galaxies as cross-calibration standard X-ray sources. Massive clusters have several advantages as calibration sources: the X-ray emission of the hot intracluster medium does not vary with time, it is bright across a broad energy band, and it has a fairly simple continuum-dominated spectrum. However, clusters are spatially extended and often contain bright, variable AGN, complicating comparison between instruments with very different imaging characteristic.  The Clusters WG meeting at the 15th IACHEC Workshop drew 25 in-person attendees, or 40\% of the Workshop attendees, indicating broad interest from the X-ray calibration community. Due to the importance of cluster science and the use of clusters as calibration sources for several recently launched and upcoming X-ray missions (e.g., eROSITA, XRISM, and ATHENA), and considering the past activity of the group (e.g., Nevalainen et al.\ 2010, Kettula et al.\ 2013, Schellenberger et al.\ 2015), it was agreed that the WG serves a useful purpose.

The majority of the Clusters WG session was devoted to a presentation by J.\,Nevalainen on cross-calibration of the three XMM-Newton/EPIC instruments using galaxy clusters. The details of this study are presented elsewhere (Nevalainen \& Molendi 2023), but in summary it provides a framework to explore broad-band effective area calibration by comparing observed cluster fluxes in quasi-independent energy bins between different instruments. The results from a sample of 27 clusters show good agreement among the EPIC instruments below 4.5 keV, but increasing disagreement above this energy and substantial cluster-to-cluster scatter. The presentation generated active discussion among the WG members.

This work serves as a useful case study for the Multi-Mission Study (MMS), a project aiming to compare X-ray spectroscopic results of a sample of clusters obtained with current and past X-ray missions. This project has been ongoing for several years, and we have identified representatives among the WG membership for XMM-Newton EPIC MOS and pn, Chandra ACIS, Suzaku XIS, Swift XRT, NuSTAR, HXMT, AstroSAT, NICER, and ROSAT. The representative for each instrument is tasked with gathering existing data spanning our cluster sample, applying the most recent calibration, extracting spectra and responses, and providing these to the WG chair for MMS cross analysis. This data will also be provided to the CalStats WG for inclusion in the Concordance effort. We expect to select a small number of clusters for an initial comparison, drawing from the HiFLUGCS sample (Reiprich \& B\"ohringer 2002) hot ($kT>6$ keV), nearby ($z<0.1$) systems with at least 100,000 counts accumulated in the central 6$^\prime$ in each instrument, observed no more than 3$^{\prime}$ off-axis. This study will form the bulk of the WG work during the foreseeable future.

\subsection{Detectors and Background}\label{s:det_bkg}

The Detectors and Background WG had one well attended session. The Detectors WG provides a forum for cross-mission discussion and comparison of detector-specific modeling and calibration issues, while the Background WG provides the same for measuring and modeling instrument backgrounds in the spatial, spectral and temporal dimensions.  Attendees represented many past, current, and future X-ray missions, including \athena, \chandra, \ep, \hitomi, \nustar, \suzaku, and \xmm.  As existing missions go deeper, and planned missions get more ambitious, understanding and modeling background and detector response is all that much more important.

The working group session had four talks, plus an extended discussion.  Bev LaMarr discussed work done at MIT on depletion studies of high thickness to pixel size ratio silicon devices which will be relevant for many future missions.  Ivan Valtchanov presented \xmm\ EPIC-pn studies of spatial CTI correction at high energies and how to include those in analysis tools.  Catherine Grant reviewed work on \chandra\ ACIS background variability, which was followed by Terry Gaetz's report on temperature-dependent ACIS response calibration.  Finally, we had a discussion of features seen in spectral fitting around the Silicon-K edge and the previous solutions used by \suzaku\ XIS and \chandra\ ACIS. Slides from the presentations are available on the IACHEC web page.

\subsection{Thermal SNRs}\label{s:t_snr}\label{s:n132d}

The thermal supernova remnants (SNRs) WG aims to use the mostly time-invariant, line-rich spectra of 1E 0102.2-7219 (E0102), Cas A, and N132D to improve the response models of the various instruments (gain, CTI correction, QE, spectral redistribution function, etc) and to compare the absolute effective areas of the instruments at the energies of the bright line complexes. Our efforts focus on developing standard models that can be used by the various teams to meaningfully compare their results.  The intention is that future missions, such as the \xrism\ mission scheduled for launch in August 2023, will be able to use these standard models to improve their calibrations. The group met remotely once in March 2023 before the IACHEC meeting itself to coordinate our efforts and prepare for the meeting.  The Thermal SNRs working group had one session on Wednesday April 26th of the IACHEC meeting.


 N132D offers some complimentary differences compared to E0102 that make it attractive as a standard candle.  The N132D spectrum contains significant amount of Fe emission, whereas the E0102 spectrum is almost completely devoid of Fe, and the N132D spectrum has significant flux at energies above 2.0 keV including an Fe-K line complex. N132D is slightly larger than E0102 which might reduce the effects of pileup in the spectrum and N132D has a higher overall flux. Martin Stuhlinger (ESAC) presented a detailed description of the development of the N132D empirical model based on the RGS data from \xmm\ that covers the bandpass from 0.3 to 1.5 keV.
He compared this phenomenological model to the physical model that was published in Suzuki~et~al. 2020.  Both models do a reasonable job at representing the emission from the bright lines or line complexes, with some notable exceptions for the \ion{Ne}{X}~Ly$\beta$ line for the empirical model and \ion{Ne}{X}~Ly$\alpha$ for the physical model. There are issues with the continuum models that require investigation for both models.  The empirical model underestimates the continuum below the  \ion{O}{VII}~He$\alpha$ triplet and above the \ion{Ne}{X}~Ly$\alpha$ line. The physical model overestimates the continuum below the \ion{O}{VII}~He$\alpha$ triplet and underestimates the continuum above the \ion{O}{VIII}~Ly$\alpha$ line.  Adam Foster (SAO) presented an analysis of the Fe-K complex based on MOS and pn data from \xmm.  He selected a total of 353/785/813 ks of pn/MOS1/MOS2 data after filtering on times of low background and complete coverage of the remnant within the spatial window used for the observation.  The focus of this analysis is to model the high energy (E~$> 3.0$ keV) to complement the RGS analysis, therefore, the Suzuki model was adopted for the low energy part of the spectrum and an additional thermal component was added to represent the spectrum above 3.0 keV. The fits can not distinguish between an ionizing, equilibrium, or recombining plasma.  The equilibrium fits prefers a temperature of kT=4.5 keV and an Fe abundance which is approximately seven times solar.  We expect to add this equilibrium component to the standard IACHEC model for N132D but we eagerly await the \xrism\ observation which should significantly improve the characterization of the Fe-K complex.

Nick Durham (SAO) presented his results on using Cas A to derive a gain calibration for ACIS on \chandra. Given that the External Calibration Source (ECS) on ACIS is a radioactive source that has decayed to less than 1.0\% of its flux at launch, the Chandra X-ray Center (CXC) calibration team has been using Cas~A to characterize changes in the CTI and gain. Cas~A has the highest flux of any Galactic supernova remnant in the \chandra\ bandpass but it has a complicated spectrum with significant spatial variations across the remnant. One source of these spectral complexities is the bulk velocity structure of the ejecta in  Cas~A, with the ejecta varying from -4,000 to +6,000  $ {\mathrm {km\, s^{-1}}}$ in velocity.  In addition, temporal variations are observed in Cas~A with \chandra\ due to the expansion of the remnant and time-dependent ionization effects. Therefore, Durham has attempted to account for the spatial variations in the spectra by developing models specific for smaller regions of the remnant and then comparing only those regions with each other over time. G\"unther (MIT) has suggested a principal component analysis that could be used with the fit results from the Cas~A spectra to determine how the gain varies as a function of time and position on the ACIS CCDs.

\begin{figure}[ht]
    \centering
    \includegraphics[width=0.95 \textwidth]{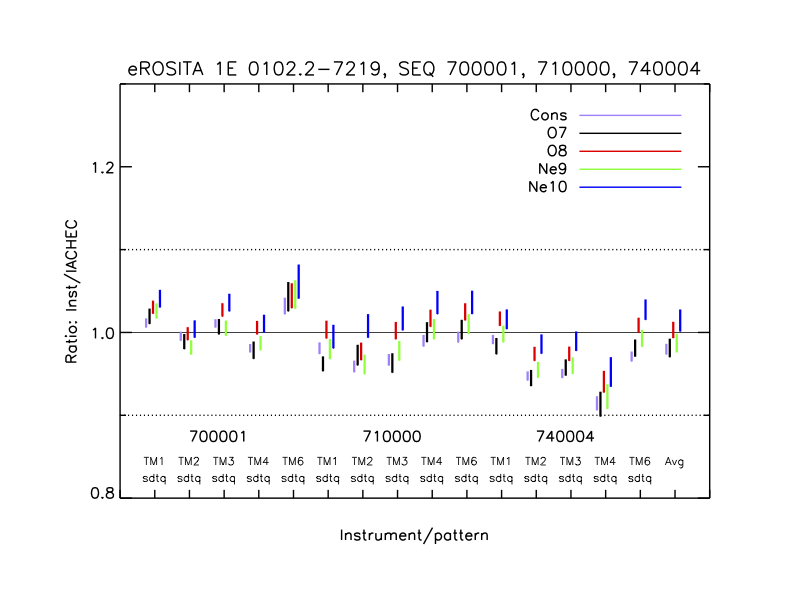}
    \caption{\ion{O}{VII}~He$\alpha$, \ion{O}{VIII}~Ly$\alpha$,
     \ion{Ne}{IX}~He$\alpha$ and \ion{Ne}{X}~Ly$\alpha$
     line normalizations from the \erosita\ E0102 fits for cameras TM1, TM2, TM3, TM4, and TM6 for all pattern events 
    (sdtq) for the 2019, 2020, \& 2021 data. }
    \label{figure:e0102}
\end{figure}


The IACHEC standard model for 1E 102.2-7219 (E0102) \footnote{\url{https://wikis.mit.edu/confluence/display/iachec/Thermal+SNR}} is routinely used by several groups for calibration purposes.  The model is described in detail in our IACHEC paper Plucinsky et al. 2017.  Paul Plucinsky (SAO) presented the latest results from E0102 to monitor the contamination buildup on ACIS using the latest N0015 contamination model. The flux at the \ion{O}{VII}~He$\alpha$ triplet has gotten so low that it is no longer useful for verifying the contamination model.  The \ion{O}{VIII}~Ly$\alpha$ line fluxes are consistent in time over the mission within the uncertainties but the uncertainties are large.  The \ion{Ne}{IX}~He$\alpha$ triplet and \ion{Ne}{X}~Ly$\alpha$ fluxes provide reliable results even at this point in the mission and show that the N0015 contamination model provides consistent fluxes over the course of the mission.  Konrad Dennerl (MPE) and Plucinsky presented fits from E0102 to the \erosita\ data.  Dennerl showed a summary of the fits to the spectra from TM1, TM2, TM3, TM4, and TM6 cameras from 2019, 2020, and 2021 for single pixel events and all pattern events. These E0102 fit results were used to improve the gain calibration of the \erosita\ CCDs at energies below 1.0 keV.  After the improved gain calibration, the E0102 fits show a gain offset that varies between -6 eV to + 3 eV.  Plucinsky compared the line normalizations for the  \ion{O}{VII}~He$\alpha$, \ion{O}{VIII}~Ly$\alpha$,
\ion{Ne}{IX}~He$\alpha$ and \ion{Ne}{X}~Ly$\alpha$ line complexes to the values in the IACHEC standard model.  The results are shown in Figure~\ref{figure:e0102}. All {\em eROSITA} line normalizations are within +/- 10\% of the IACHEC values, but there appear to be some significant variations in time that warrant further investigation.

\subsection{Timing}\label{s:timing}

The Timing WG aims to provide a forum for in-orbit and on-ground timing calibrations of X-ray missions, focusing on their timing systems, calibration methods, issues, and lessons learned. The WG also aims to coordinate simultaneous observations for timing calibrations with multi-X-ray missions and/or radio observatories.

The WG has three major goals: {\bf a)} sharing information on timing calibration methods and protocol, lessons learned to enhance timing capability, {\bf b)} performing the cross-calibration on the timing performance among multiple missions, analyzing systematically the archive data and/or triggering coordinated observations and discussion on the calibration plan for the near future missions, and {\bf c)} studying in detail the effects on the timing products (i.e., light curve, power spectrum, etc) by the detector’s behaviors, such as the dead time, the good-time-interval selection, grade selection, etc.

At the 15th IACHEC meeting, we made the following progress for goal {\bf a)}: We had three presentations on the timing calibration status of X-ray missions: “\xrism\ Timing System Design and Timing Accuracy” by Yukikatsu Terada, “Verification of \xrism\ Timing System Using Thermal-vacuum Test Data” by Megumi Shidatsu, and “\hxmt\ in-orbit timing calibration” by Youli Tuo. The first two presentations showed the detailed design of the timing system and the latest status of ground timing calibration of the \xrism\ satellite. The third showed the timing system of the \hxmt\ satellite and summarized the current status and issues of the on-orbit time calibration.

Together, goals {\bf b)} and {\bf c)} cover three projects, as listed in Table \ref{table:WGoverview}; {\bf i)} summary of Timing performance and calibration status across missions; {\bf ii)} systematic timing cross-calibration with Crab archive data; and {\bf iii)} planning timing cross-calibration campaign.

\subsubsection{Summary of timing calibration and performance}

On project {\bf i)}, the first version of the summary table of the timing calibration and performance is completed for 20 instruments on 12 missions (\textit{RXTE}, \chandra, \xmm, \integral, \swift, \suzaku, \nustar, \fermi, \astrosat, \hitomi, \nicer, and \xrism), which is shown in \url{https://wikis.mit.edu/confluence/display/iachec/Timing} with columns below.

\begin{enumerate}
\item Science Requirement Absolute Time (Requirement \& Goal)
\item Timing System Design (GPS yes/no, Clock Stability)
\item Timing Calibration Status (Offset from the Reference,	Deviation, Reference Time, Notes)
\item In-orbit Timing Calibration Targets
\item Reported Issues
\item Reference
\end{enumerate}

Before the 15th IACHEC meeting, we updated the definition of the 3) the "Timing Calibration Status". The updated table has timing ``offset", ``reference time", and the ``deviation". The definitions of these columns are as follows: the ``Offset" is defined as the average value of the difference time from the reference time, which is defined in the ``Reference Time" column, and the ``Deviation" represents the fluctuations in ``Offset" time that have been observed many times. Here, ``Reference Time" are categorized as:

\begin{description}
    \item[TAI] Engineering value. The reference time is the International Atomic Time (TAI).
    \item[Radio (observatory\_name)] Radio ephemeris of Crab main pulse by observatory\_name (ex., Jodrell Bank radio observatory).
    \item[Xray (instrument\_name, $N$)]  X-ray ephemeris of Crab main pulse by the instrument\_name (ex., NICER), assuming the timing reference from radio main pulse is $N \mu$sec (ex. 350 $\mu$sec).
    \item[Gamma (observatory\_name, $N$)]  Gamma-ray ephemeris of Crab main pulse by observatory\_name (ex., Fermi LAT), assuming the timing reference from radio main pulse is $N \mu$sec (ex. 350 $\mu$sec).
    \item[Other] Definition is defined in ``Note" column
\end{description}

At the 15th IACHEC meeting, we decided to try to add \erosita, \xmm\ EPIC-MOS, and \integral\ SPI, on the table.
In addition, to make this table mode visible, we added the ``Current Activities" page on the IACHEC web page to show the highlights of the outputs from the projects.

\subsubsection{Systematic timing study with Crab archive data}

This project is responsible for coordinating the systematic timing-cross-calibration among instruments using the archive data of Crab and see the ephemeris among instruments. There have been no major updates on this activity since the last IACHEC virtual workshop. 

\subsubsection{Planning simultaneous timing cross-calibration with \xrism\ and \nicer\ in 2024}

At the 15th IACHEC meeting, the \xrism\ members propose the cross-timing calibration between \xrism\ Resolve and \nicer\ using Crab during the epoch from December 2023 to June 2024, assuming the launch of \xrism\ in August 2023. From the point of view of the visibility, this campaign can start from February 2024.

At the meeting, the following missions were confirmed to participate in this campaign and assigned personnel for operational coordination and analysis; \xrism\ (Yukikatsu Terada), \nicer\ (Craig Markwardt), Radio in Japan (Teru Enoto; to be confirmed), \hxmt\ (Xiaobo Li), and \ep\ (Juan Zhang; if possible).

\section*{References\footnote{see \url{https://iachec.org/papers} for a complete list of IACHEC papers}}


\noindent
Bonamente, M.\ 2020, Journal of Applied Statistics, 47, 2044, arXiv:1912.05444\\
\noindent
Bonamente, M.\ 2022, MNRAS, arXiv:2302.04011\\
\noindent
Chen, Y., et al.\ 2019, J.Am.Stat.Assoc., 114:527, 1018, doi.org/10.1080/01621459.2018.1528978\\ 
\noindent
Kettula, K., Nevalainen, J., \& Miller, E.D.\ 2013, A\&A, 552, 47\\
\noindent
Marshall, H. 2021, AJ, 162, 134\\
\noindent
Marshall, H. 2024, ApJ, 964, 88\\
\noindent
Marshall, H., Chen, Y., Drake, J.J., Guainazzi, M., Kashyap, V.L., Meng, X.-L., Plucinsky, P.P., Ratzlaff, P., van Dyk, D.A., \& Wang, X.\ 2021, AJ, 162, 254\\
\noindent
Nevalainen, J., David, L., \& Guainazzi, M.\ 2010, A\&A, 523, A22\\
\noindent
Nevalainen, J. \& Molendi, S.\ 2023, A\&A, 676, A142\\
\noindent
Plucinsky, P., et al.\ 2017, A\&A, 597, A35 \\
\noindent
Reiprich, T.H. \& B\"ohringer, H.\ 2002, ApJ, 567, 2, 716\\
\noindent
Schellenberger, G., et al.\ 2015, A\&A, 575, 30\\
\noindent
Schneider, B., et al.\ 2022, Experimental Astronomy, arXiv:2212.09863 \\
\noindent
Silvestri, S., and the IXPE Collaboration 2023, Nuclear Instruments and Methods in Physics Research A, 1048, 167938\\
\noindent
Suzuki, H., et al.\ 2020, ApJ. 900, 39\\
\noindent
Sweere, S., Valtchanov, I., Lieu, M., Vojtekova, A., Verdugo, E., Santos-Lleo, M., Pacaud, F., Briassouli, A., \* C\'{a}mpora P\'{e}rez, 2022, MNRAS, 517, 5044\\
\noindent
Zhang, X., Algeri, S., Kashyap, V.L., \& Karovska, M., 2023, MNRAS, 521, 969\\

\end{document}